\documentclass[10pt,letterpaper,twocolumn]{article}

\usepackage[letterpaper,
  top=0.98in,
  bottom=0.98in,
  left=0.98in,
  right=0.98in,
  headheight=22pt,
  headsep=8pt,
  footskip=24pt]{geometry}

\usepackage[T1]{fontenc}
\usepackage{mathptmx} 
\usepackage{microtype}
\usepackage{graphicx}
\usepackage{booktabs}
\usepackage{array}
\usepackage{cuted}
\usepackage{amsmath}
\usepackage{enumitem}
\usepackage{titlesec}
\usepackage{fancyhdr}
\usepackage{lastpage}
\usepackage{caption}
\usepackage{url}
\usepackage[hidelinks]{hyperref}

\newcommand{\IACPaperCode}{IAC-26,A4,1,1,x113931}
\newcommand{\IACTitle}{The First Decade of Breakthrough Listen}
\newcommand{\IACConference}{77\textsuperscript{th} International Astronautical Congress (IAC 2026), Antalya, T\"urkiye, 5--9 Oct 2026.}
\newcommand{\IACCopyright}{Copyright \textcopyright 2026 by David MacMahon. Published by the IAF, with permission and
released to the IAF to publish in all forms.}

\titleformat{\section}
  {\normalfont\bfseries\fontsize{10}{11}\selectfont}
  {\thesection.}{0.45em}{}
\titlespacing*{\section}{0pt}{9pt}{1pt}

\titleformat{\subsection}
  {\normalfont\itshape\fontsize{10}{11}\selectfont}
  {\thesubsection}{0.45em}{}
\titlespacing*{\subsection}{0pt}{7pt}{1pt}

\titleformat{\subsubsection}
  {\normalfont\itshape\fontsize{10}{11}\selectfont}
  {\thesubsubsection}{0.45em}{}
\titlespacing*{\subsubsection}{0pt}{6pt}{1pt}

\newcommand{\iacunnumbered}[1]{%
  \par\vspace{7pt}%
  \noindent\textbf{#1}\par\vspace{1pt}}

\setlist{nosep,leftmargin=*}

\begin{document}

\twocolumn[{%
\begin{minipage}{\textwidth}
\begin{center}
  \fontsize{11}{12}\selectfont
  \IACPaperCode\

  \vspace{8pt}

  {\bfseries\fontsize{12}{14}\selectfont \IACTitle}

  \vspace{8pt}

  {\bfseries\fontsize{11}{13}\selectfont
  David H. E. MacMahon\textsuperscript{a,b}\\
  Daniel J. Czech\textsuperscript{c}}
\end{center}

\vspace{5pt}

\noindent\textsuperscript{a}\textit{University of California at Berkeley, Berkeley, CA, USA}\par
\noindent\textsuperscript{b}\textit{Berkeley SETI Research Center, Berkeley, CA, USA}\par
\noindent\textsuperscript{c}\textit{Breakthrough Listen, Department of Physics, University of Oxford, Oxford, UK}\par
\noindent\textit{Corresponding author: davidm@berkeley.edu}\par

\vspace{11pt}

\begin{center}
\textbf{Abstract}
\end{center}
\vspace{-4pt}

\noindent
Breakthrough Listen (BL), launched in July 2015 with a 10-year, \$100M commitment from the Breakthrough Prize Foundation,
is the most comprehensive search for technosignatures ever conducted. This paper reviews the program's first decade
from the facilities and instrumentation perspective: the digital backends and data recorders, the signal-processing
pipelines, the data archive, and the observatory partnerships that made the science possible. We organize the decade into
three architectural eras. In Generation 1 (2016--2019), BL deployed packetized raw-voltage wideband recorders on the
100-m Green Bank Telescope and the 64-m Parkes telescope, enabling targeted surveys of thousands of nearby stars across
$\sim$0.7--8 GHz. In a network-expansion phase (2020--2023), BL extended its instrumentation to partner facilities including the
Sardinia Radio Telescope and international LOFAR stations, and developed verification frameworks
that proved essential when the program's highest-profile signal candidates emerged. Since 2023, BL has operated in
a fully commensal era: Ethernet-based, commodity-GPU backends riding on routine observatory operations: COSMIC at
the Karl G. Jansky VLA, a revived Allen Telescope Array with deployed real-time deep-learning pipelines, and an autonomous
multi-beam commensal survey at MeerKAT (BLUSE; $>$1.2 million pointings since mid-2022). Over the decade, capturing data
at ever higher rates forced a systematic approach to compression and archiving; the storage estate grew to $\sim$42 PB of
capacity ($\sim$31 PB in use) across seven systems on four continents by March 2026. We conclude with engineering lessons
(commodity hardware over bespoke hardware, open data as a first-class requirement, commensal science as backend validation,
and verification frameworks built before they are needed) and with the outlook for SKA-era commensal technosignature science.

\vspace{2pt}

\vspace{8pt}
\end{minipage}
}]

\section{Introduction}

\subsection{The Breakthrough Listen program: origin, scale, and goals}

On July 20, 2015, at the Royal Society in London, Stephen Hawking, Lord Martin Rees, Ann Druyan, and Frank Drake announced
Breakthrough Listen: a 10-year, \$100 million astronomical program to search for evidence of technological life beyond
Earth. Managed by the Breakthrough Initiatives on behalf of the Breakthrough Prize Foundation (Worden et al. 2017),
the program defined from the outset an observing portfolio of unusual breadth: sensitive targeted searches of the
nearest several thousand stars in both radio and optical; a shallower ``million-star'' survey to be conducted
commensally on radio arrays; surveys of the Galactic plane and of nearby galaxies; and development toward
all-sky monitoring at radio and optical wavelengths (Worden et al. 2017).

Observationally, the program began with the Lick Observatory 2.4-m Automated Planet Finder (APF) commencing
optical SETI observations of 200 nearby stars in December 2015.  That same month a new data recorder saw first light on
the 100-m Robert C. Byrd Green Bank Telescope (GBT).  In October 2016 the 64-m radio telescope Murriyang in Parkes,
Australia, took its first observations for the program.  These three facilities, announced
at the launch, marked the initial phase of the Breakthrough Listen program. (Worden et al. 2017).

BL inherited a specific instrumentation culture from its Berkeley SETI roots (SERENDIP commensal backends and the
volunteer-computing experiment SETI@home; Korpela et al. 2025; Anderson et al. 2025) and from long-standing
collaboration with NRAO and observatory engineering staff. The program's founding team explicitly framed its search
strategy in terms inherited from the SETI 2020 roadmap and decades of prior practice (Worden et al. 2017).

The project's scope and ambition were unprecedented.  The first few months were
spent developing the core instrumentation that would form the foundation of BL's
data acquisition model for the years that followed, scaling as technology
progressed.  Interfacing with components of the observatories' digital
instrumentation allowed BL to leverage the already operational infrastructure,
thereby saving development and testing time and allowing the team to focus on the
data capture and analysis software.

\subsection{Roadmap}

This history draws on the published BL literature, conference
presentations, and the authors' first-hand experience. Section~\ref{sec:design} describes the design philosophy that shaped all BL
hardware and software. Section~\ref{sec:gen1} covers Generation 1
wideband recorders (2016--2019). Section~\ref{sec:network} covers the network-expansion years
(2020 to 2023). Section~\ref{sec:commensal} describes the commensal era (2023--present). Section~\ref{sec:archive} treats the data archive. Section~\ref{sec:signalproc} traces signal-processing and RFI
engineering across all eras. Section~\ref{sec:validation} discusses science and spacecraft for system validation. Section~\ref{sec:future} looks to future
facilities, and Section~\ref{sec:lessons} considers lessons learned.

\section{Design Philosophy and Program Architecture}\label{sec:design}

\subsection{Open data as a design requirement}

BL committed from the start to public release of its data products (Worden et al. 2017). That commitment had concrete
engineering consequences. A single radio telescope with a wideband raw-voltage backend can generate of order a
petabyte per day of raw-voltage data. We therefore adopted the approach of preserving all the reduced data and only a
curated raw-voltage subset at full fidelity (of order 1 PB/yr), with both types of data publicly available.

The program standardized its file formats early: SIGPROC filterbank products with documented metadata (eventually migrating to HDF5), plus
raw-voltage GUPPI-style formats, were codified in the public data-release and archiving paper (Lebofsky et al. 2019),
building on format and compute-framework work presented as early as 2016 (Price 2016). BL also engaged the open-source
signal-processing community directly, presenting its pipelines at GNU Radio conferences (Croft et al. 2019; Croft and
Pollak 2020). In the commensal era, the per-pointing ``hits'' and ``stamps'' files generated by BLUSE and COSMIC
(Sections~\ref{sec:bluse} and~\ref{sec:cosmic}) joined the public data stream.

\subsection{Commensality and partnership models}

The million-star commensal survey was a founding goal of Breakthrough Listen.  The 2017 program description already
anticipated piggyback technosignature observations on next-generation arrays, naming MeerKAT, the Murchison Widefield
Array, and FAST as future platforms (Worden et al. 2017), and commensal SETI in the SKA era was presented in 2016
(Gajjar 2016). In practice, commensality took over half a decade to become the program's dominant mode (Section~\ref{sec:commensal}), passing first
through observer-based partnerships in which BL scientists brought their own backends to partner observatories (NRAO,
CSIRO, INAF, and LOFAR) and were allocated observing time in exchange for data sharing, commensal capability, and in some
cases funding.

\subsection{Target-selection science as the driver of observing campaigns}

BL's primary (as opposed to commensal) observing campaigns derived from explicit target lists.  In addition to the galactic plane survey, a nearby-star
and nearby-galaxy sample was selected based on
sensitivity and diversity (Isaacson et al. 2017).  This was later joined by the Exotica Catalog, comprising ``one of everything'' from intermediate-mass black holes to
WDs and ultracool dwarfs (Lacki et al. 2021). Frequency coverage, cadence strategy, and data-volume budgets all trace
back to these lists. The commensal era added target catalogues of its own: the BLUSE survey draws on a Gaia-derived
list of $\sim$32 million stars (Section~\ref{sec:bluse}), complemented by a ``semi-Exotica'' catalogue that adapts the Exotica
approach to commensal observing.

\subsection{Search-parameter formalism that shaped pipelines}\label{sec:searchparams}

BL pipelines were designed around physically motivated search parameters rather than convenience. The choice of maximum
drift rate, i.e., how fast a narrowband signal can slide across the spectrometer's channel grid, was derived from the
dynamics of transmitters on rotating bodies (Sheikh et al. 2019). Observing strategy (on/off cadencing for
terrestrial-signal rejection, beam/pointing coincidence) was quantified to maximize detection rate per unit telescope
time (Enriquez et al. 2017; Houston et al. 2021).

\section{Generation 1: Dedicated Wideband Recordings (2016--2019)}\label{sec:gen1}

\subsection{The Green Bank Telescope wideband backend}

BL's first major instrument was a wideband data recorder for the GBT.  This provided raw-voltage capture over commodity
Ethernet of up to 10 GHz of instantaneous bandwidth comprising 1.5 GHz bands spanning from 0.7--12 GHz (across L, S, C, and X
bands).  Data were written to a GPU-equipped processing cluster (Worden et al. 2017; MacMahon et al. 2018). The design favored
commodity PCI-Express servers, 10 GbE packet capture, and channelization on GPUs.  This bet on off-the-shelf hardware benefited every
later BL backend.

The same backend produced the program's first survey results: a 1.1--1.9 GHz search of 692 nearby stars (Enriquez et al.
2017), an extension of the 1.10--3.45 GHz survey covering 1,327 stars (Price et al. 2020), and the highest-frequency detection of
the repeater FRB 121102 at 4--8 GHz, a demonstration that the SETI instrument was simultaneously a capable transient
machine (Gajjar et al. 2018). The two GBT surveys alone logged a combined total of approximately 1,445 hours of on-sky
time (Enriquez et al. 2017; Price et al. 2020).

Raw-voltage recording rates and post-reduction data volumes were separated by orders of magnitude. Recording ran at
6 Gbps per node, with eight nodes per bank and up to four banks in use at once, for aggregate capture rates of up to
192 Gbps (24 GB/s, or roughly 86 TB/hr). The surveys in turn produced 180 TB (Enriquez et al. 2017) and 142 TB
(Price et al. 2020) of reduced filterbank products, and in most cases the raw voltage data files were deleted once those products
were generated, with only select targets, such as FRB 121102, retained in raw form for further analysis (Lebofsky et al. 2019).

\subsection{Parkes 64-m wideband digital instrumentation}

In the southern hemisphere, BL deployed two early digitizer systems on Parkes covering 0.7--4 GHz (Price et al. 2018).
The first, a ROACH2-based recorder very similar to the GBT data recorder, served the single-pixel receivers; the
second, built from SNAP boards, handled the 21-cm multibeam receiver used for the Galactic plane survey. Both shared
the same compute cluster and supported similar data products to the Green Bank system, including high spectral
resolution data for narrowband SETI. Parkes later moved
to a direct-to-filterbank recording mode, later replaced by direct-to-HDF5, under which raw voltages were not written
to disk by default but could still be captured if and when needed. This instrumentation fed targeted surveys and,
later, multibeam Galactic-plane survey concepts (Solaiyappan et al. 2024), and its data products underpinned
southern-hemisphere searches such as the TESS-eclipsing-exoplanet survey (Barrett et al. 2025; Choza et al. 2024).

\subsection{Optical facilities in Gen 1}

Optical SETI led the program: APF observations began in December 2015 (Worden et al. 2017), motivated by the economics
of laser communication, including the beam-director scaling arguments of the Breakthrough Starshot initiative (launched
by the Breakthrough Initiatives in April 2016). The first era's optical program centered on high-resolution
spectroscopic searches for laser line emission on Keck/HIRES and APF (Tellis and Marcy 2017; Lipman et al. 2019;
Isaacson et al. 2019), later consolidated into an automated laser-search pipeline on the APF (Zuckerman et al. 2023).

\subsection{Compute and software stack of Gen 1}

Gen 1 established the program's software stack: GPU-based coarse channelization of raw voltages into filterbank
products on the recorder clusters, and dedoppler search (turboSETI-style) over those products. First-generation machine learning arrived on
two fronts: a CNN for FRB 121102 pulse detection and periodicity on the transient side (Zhang et al. 2018a), and
self-supervised anomaly detection for narrowband SETI (Zhang et al. 2018b).

\section{Growing the Network: Additional Partnerships (2020--2023)}\label{sec:network}

\subsection{European partnerships}

BL extended to the Sardinia Radio Telescope (SRT) through a long-standing collaboration between INAF and Berkeley,
using SRT for both commensal FRB work (including the lowest-frequency detection of the periodic repeater FRB 180916 at
328 MHz; Pilia et al. 2020) and the first high-frequency technosignature survey program at the facility (Manunza et al. 2025).

BL's low-frequency extension used two international LOFAR stations in Europe (Ireland and Sweden) in a simultaneous
dual-site configuration. Beamformed baseband voltages were recorded at both stations, barycentric corrections were
applied in post-processing, and dual-station coincidence detection provided the terrestrial-signal rejection strategy
(Johnson et al. 2023).

\subsection{Other optical and VLBI projects}

Two further modalities rounded out the network years and are noted here for completeness: a pulsed optical
technosignature search with the VERITAS imaging atmospheric-Cherenkov telescope, combining 30 hours of dedicated and
110 hours of archival observations of Breakthrough Listen targets (Acharyya et al. 2023), and an interferometric SETI
demonstration toward Kepler-111 b with the European VLBI Network at 21 cm, combining a 3-hour, 11-antenna session in
November 2019 with a 1-hour, 7-antenna follow-up in March 2022, plus supporting e-MERLIN observations (Wandia et al. 2023).

\subsection{Cross-cutting campaigns and high-profile follow-up}

The network years were stress-tested by targets of opportunity: the Restricted Earth Transit Zone survey at
3.95--8.00 GHz (Sheikh et al. 2020), the Galactic Center program (Gajjar et al. 2021), TESS targets of interest (Traas
et al. 2021; Franz et al. 2022), a GBT search during Kepler planetary transits (Sheikh et al. 2023), a
solar-gravitational-lens pointing toward Alpha Centauri (Tusay et al. 2022), and multi-wavelength follow-up of
Boyajian's Star (Wright et al. 2017).

The defining episode of this period was the Proxima Centauri signal of interest: a narrowband drift signal found in
Parkes archival pointing data (Smith et al. 2021), subjected to a purpose-built technosignature verification framework
(lookalike analysis, reobservation campaigns, and instrumental-artifact tests) and ultimately attributed to terrestrial
interference (Sheikh et al. 2021; Sheikh et al. 2021b). The verification framework developed for this episode became
the program's standard for candidate vetting (Section~\ref{sec:verification}).  This was the first signal to pass all our verification
filters at that time.  It became known as Breakthrough Listen Candidate 1 (BLC1).  It caused the BL team to reflect on
the threshold a signal would have to reach before declaring a detection of an extraterrestrial technosignature.  BL's
grounding in evidence-based science demanded that the evidence be solid.  The governing maxim was clear: extraordinary
claims require extraordinary evidence or proof (Truzzi 1978; Sagan 1979).  This led to an exciting all-hands-on-deck
re-analysis of the data to determine whether it was of human or extraterrestrial origin.  Ultimately, the detection of
signals with morphologies similar to BLC1, but at different pointings, led to the determination that it was in fact of
human origin.

\section{The Commensal Era: Streaming and Cloud-Native Instruments (2023--present)}\label{sec:commensal}

\subsection{MeerKAT: from IAC Bremen announcement to the BLUSE autonomous survey}\label{sec:bluse}

\begin{figure*}[t]
\centering
\includegraphics[width=\textwidth]{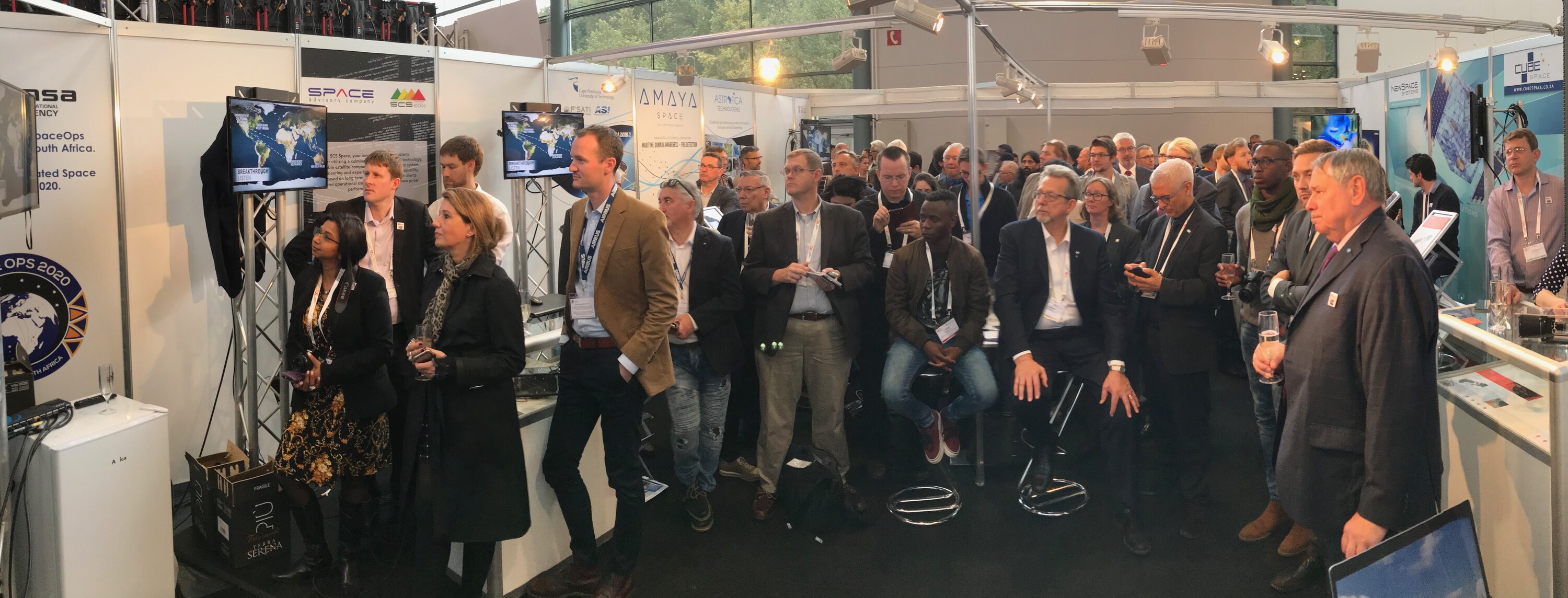}
\caption{Announcement of the Breakthrough Listen/SARAO partnership at the International Astronautical Congress in
Bremen, Germany, in 2018.  The partnership led to the BLUSE backend on the MeerKAT array (Section~\ref{sec:bluse}).}
\label{fig:bremen}
\end{figure*}

Breakthrough Listen's engagement with MeerKAT was first announced at the International Astronautical Congress in Bremen
in 2018 (Figure~\ref{fig:bremen}), and the partnership matured through a MeerKAT-specific target-selection
and observing strategy (Czech et al. 2021) and early single-dish-mode tests (Heywood et al. 2024). MeerKAT's packetized
architecture is the enabling factor: F-engine data are distributed via IPv4 multicast, allowing ``user-supplied
equipment'' (USE) backends to subscribe to the full array bandwidth in parallel with the correlator. BLUSE, Breakthrough
Listen's User Supplied Equipment system, has exploited this continuously and autonomously since mid-2022 (Czech et al.
2026). BLUSE ingests the full F-engine bandwidth (27.4 Gbps per antenna at L-band; $\sim$1.75 Tbps for the full
64-antenna array), upchannelises to $\sim$1 Hz resolution with \texttt{seticore}, synthesizes 64 coherent beams plus one
incoherent beam per 290-s primary pointing on targets drawn from a $\sim$32-million-star Gaia-derived list, and runs a
Taylor-tree dedoppler search ($\pm$4 Hz/s), saving ``hits'' and ``stamps'' files, the latter being upchannelized
raw-voltage swatches that can be used for offline re-beamforming and re-analysis. The system runs on a SARAO-hosted cluster of 16 racks in
the Karoo Array Processor Building, with coordinator/state-machine automation that has grown the survey rate five-fold:
$\sim$73,500 coherent beams in the first year, $\sim$386,200 in the second, and $\sim$436,700 in the third. In total,
$\sim$1.5 million beams were processed, of which $\sim$1.2 million pointings were viable for technosignature searching,
a sustained rate of $\sim$29,000 previously unobserved stars per month. BLUSE was validated end-to-end against a known
anthropogenic ``technosignature'': the JWST S-band telemetry downlink transiting MeerKAT's field of view (Czech et al.
2026). In parallel, the TRON project mines the interferometric visibility stream, the image-domain complement to BLUSE's
beamforming, detecting millisecond pulsars and stellar flares entirely from other programs' data (Smirnov et al. 2025a;
Smirnov et al. 2025b).

\subsection{COSMIC at the Karl G. Jansky VLA}\label{sec:cosmic}

COSMIC is the clearest expression of the commensal-era architecture: an Ethernet-based, multimode digital backend that
taps the VLA's signal path, transports packetized channelized data over the network to a GPU cluster, and runs SETI
searches alongside, and never interferes with, the observatory's science program (Tremblay et al. 2024). COSMIC's first
large survey rode the VLA Sky Survey, searching VLASS fields commensally with a documented pipeline and first results
(Tremblay et al. 2025); the tooling ecosystem around it now includes dedicated search pipelines (Steffes et al. 2024),
statistical RFI flagging (Sofair et al. 2024), and an ongoing high-frequency extension (Stiegler et al. 2025).

\subsection{Allen Telescope Array revival}

The Allen Telescope Array (ATA), refreshed with wideband feeds and CASPER-style backends with BL participation, became
a workhorse for both technosignature and transient science: characterization of the repeating FRB 20220912A (Sheikh et
al. 2024), a TRAPPIST-1 technosignature search (Tusay et al. 2024), and high-frequency wideband FRB studies (Joshi et
al. 2025). The ATA's rapid-response capability was exercised repeatedly in 2024--2025, from X-ray-binary flares (Farah
et al. 2024) and tidal-disruption events (Sfaradi et al. 2025) to the interstellar object 3I/ATLAS (Sheikh et al. 2025).

\subsection{Real-time, deployed ML}\label{sec:realtimeml}

The commensal era's most consequential software shift is the deployment of machine learning \textit{inside} the data
path rather than after it. At the ATA, a modified masked ResNet-38 model performs end-to-end FRB detection on
$\sim$100 Gb/s beamformed spectrograms \textbf{without} dedispersion or RFI masking stages, deployed on NVIDIA's
Holoscan GPU sensor-processing platform with NVIDIA co-authorship; the deployed pipeline runs 150$\times$ faster than
real-time production constraints and was validated live on Crab-pulsar giant pulses and offline on FRB 20240114A
(Ma et al. 2025). In parallel, technosignature searches are beginning to integrate with time-domain alert brokers for
target-of-opportunity follow-up (Gallay et al. 2025; Davenport et al. 2023).

\subsection{Gaia-anchored time-domain SETI}

Commensal-era searches increasingly anchor on precise astrometric and temporal priors, such as signal synchronization
strategies using Gaia DR3 and SETI-ellipsoid event timing (Nilipour et al. 2023; Davenport et al. 2022), a mode enabled
by the same streaming infrastructure.

\section{Building the Archive: Data Management at Petabyte Scale}\label{sec:archive}

\subsection{Volume engineering}

Gen-1 recorders produced raw-voltage streams at tens of Gbps. BL's response was tiered: full-fidelity capture to
on-site buffers, near-line reduction to filterbank and derived products, and curation of a raw subset for permanent
archive (Worden et al. 2017; Lebofsky et al. 2019). As of March 2026, BL-operated storage totals $\sim$42 PB of
capacity across seven systems, with $\sim$31 PB in use (Table~\ref{tab:storage}).  The data storage spans several
observatories (Green Bank, Parkes, Sardinia, MeerKAT, the ATA, and COSMIC/VLA) as well as the main data archive
currently hosted ``off prem'' at the Berkeley data center.

\begin{table}[htbp]
\caption{BL storage inventory, March 25, 2026}
\label{tab:storage}
\centering
\fontsize{9}{10}\selectfont
\begin{tabular}{@{}lrr@{}}
\toprule
System & Capacity (PB) & Used (PB) \\
\midrule
GBT (Green Bank) & 9.7 & 5.8 \\
Parkes & 3.5 & 3.1 \\
SRT (Sardinia) & 0.2 & 0.2 \\
DC (Berkeley data center) & 16.0 & 15.0 \\
COSMIC (VLA/DC) & 3.6 & 3.1 \\
MeerKAT & 7.0 & 2.7 \\
ATA & 1.8 & 1.5 \\
\midrule
\textbf{Total} & \textbf{41.8} & \textbf{31.4} \\
\bottomrule
\end{tabular}
\end{table}

Inter-site transfer from Green Bank, Parkes, and later partners to Berkeley was
itself an engineering effort.  We tried shipping boxes of disks, but ultimately found that approach to be a bit
cumbersome, both technically and logistically.  In the end, the most practical solution was often a steady stream of
data over the internet, even with only modest data egress rates from remote facilities.  We are currently exploring
the use of magnetic tape for data transport and archival storage purposes.

\subsection{The archive as an experimental facility}

The BL public archive (Lebofsky et al. 2019) has proven to be an experimental facility in its own right. Because
raw-voltage retention preserved the option to reprocess with better algorithms, including deep learning, archive-scale
reprocessing years after acquisition now supports novel searches that were impossible at the time of observation, for
example unsupervised anomaly-detection campaigns over the GBT narrowband archive (Painter et al. 2025; Jacobson-Bell
et al. 2025). Open tooling compounds this value: the \texttt{setigen} framework standardized injection/recovery testing
for search pipelines (Brzycki et al. 2022), enabling quantitative end-to-end validation of both classical and ML-based
searches against archived data. The resulting lesson is that archiving converted a storage cost into a scientific
multiplier. That is why the archive, rather than any single telescope, may be BL's most durable instrumentation legacy.

\section{Signal-Processing and RFI Engineering Across Eras}\label{sec:signalproc}

\subsection{Detection \texorpdfstring{$\rightarrow$}{->} verification pipelines}\label{sec:verification}

BL's search pipelines pair physically motivated detection parameters (Section~\ref{sec:searchparams}) with a structured verification
discipline: on/off cadencing and beam/pointing coincidence at acquisition (Enriquez et al. 2017),
drift-rate-consistent dedoppler search (Sheikh et al. 2019), and, since the BLC1 episode, a formal framework covering
signal localization, RFI-lookalike databases, reobservation, and instrumental tests before any candidate is claimed
(Sheikh et al. 2021). The framework was exercised again on the Wow! signal's field in a modern reobservation campaign
(Perez et al. 2022).

\subsection{Machine learning evolution}

BL's ML arc runs from 2018-era supervised/self-supervised prototypes for narrowband anomaly detection and FRB
periodicity (Zhang et al. 2018a; Zhang et al. 2018b), through CNN-scale technosignature searches of 820 nearby stars
(Ma et al. 2023) and deep reverse-spectrogram search (Ma et al. 2024), to 2025 production deployments: unsupervised
anomaly detection and RFI classification on GBT archive data (Pardo et al. 2025; Jacobson-Bell et al. 2025) and the
Holoscan-deployed real-time pipeline of Section~\ref{sec:realtimeml}.

\subsection{Compute economics}

Search cost scales with drift-rate range, channel count, and cadence; BL has attacked the constant factors directly,
for example through reduced-resolution beamforming to cut the computational cost of pulsar and technosignature surveys
on arrays (Price et al. 2024). At the other end of the trade, BLUSE demonstrates the cost of full generality:
$\sim$1 Hz channelization of MeerKAT's entire $\sim$1.75-Tbps array feed (Czech et al. 2026), feasible only because
multicast Ethernet delivers F-engine data to commodity AMD/GPU processing nodes at low marginal cost.

\subsection{Injection/recovery and synthetic data}

Pipeline claims are validated against synthetic injections: \texttt{setigen} provides the standard simulation framework
for radio technosignatures (Brzycki et al. 2022), supporting localization studies on noisy synthetic spectrograms
(Brzycki et al. 2020) and the training sets behind the ML systems above.

\subsection{The RFI problem, quantified}

RFI is the governing background for all radio SETI. BL has characterized and mitigated satellite interference explicitly
(Murphy et al. 2019), and the satellite-constellation era has raised the stakes for the coincidence- and
statistics-based rejection strategies described above.

\section{Science and Spacecraft for System Validation}\label{sec:validation}

The BL backends were validated primarily through the observation of astrophysical sources and a handful of human-made satellites and probes.

\subsection{Spacecraft telemetry}\label{sec:spacecraft}

\begin{figure*}[t]
\centering
\includegraphics[width=0.48\textwidth]{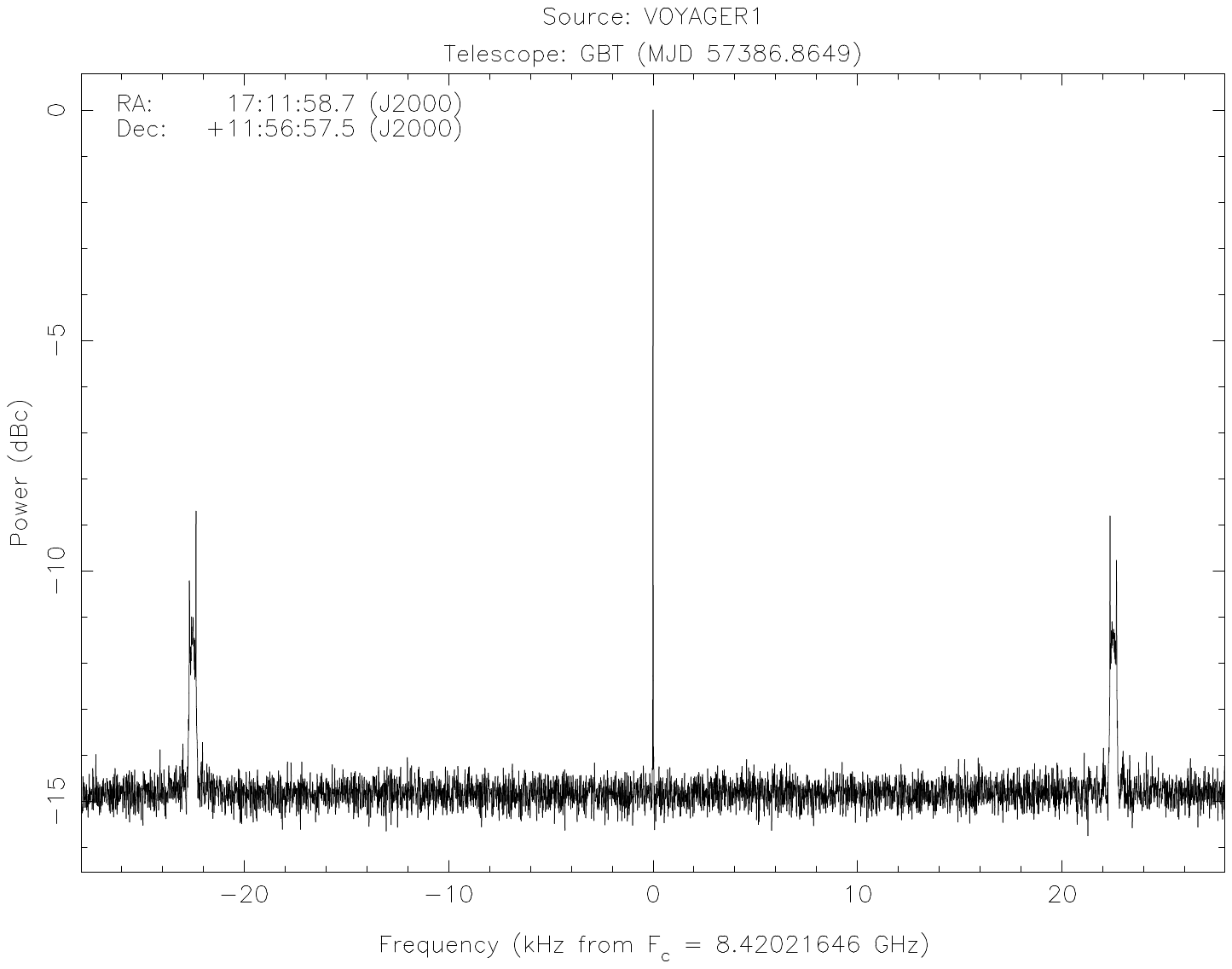}\hfill
\includegraphics[width=0.48\textwidth]{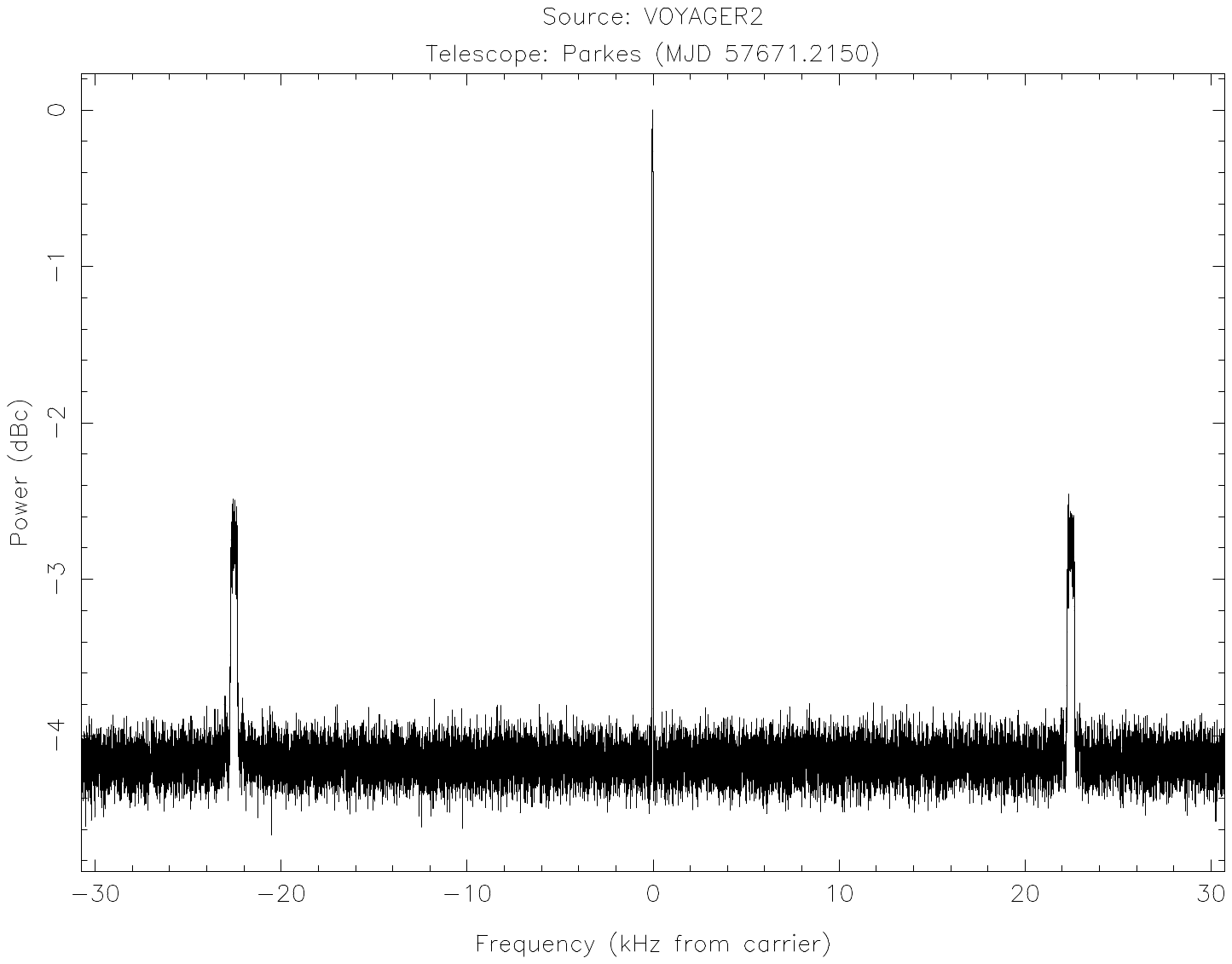}
\caption{High-resolution spectra showing the main carrier and sidebands of the Voyager 1 (left) and Voyager 2 (right)
telemetry downlinks as produced by the Breakthrough Listen systems at the Green Bank Telescope and Parkes,
respectively.  The Voyager spacecraft are the farthest human-made objects from Earth, making these faint signals of
known origin a suitable end-to-end test of the BL signal chains.}
\label{fig:voyager}
\end{figure*}

The Voyager spacecraft are the farthest human-made objects from Earth, making their
faint telemetry carriers a suitable test of the ability to detect extraterrestrial technosignatures.  The carriers of
Voyager 1 and Voyager 2 were detected with the BL systems at the Green Bank Telescope and Parkes, respectively
(Figure~\ref{fig:voyager}), exercising the full signal chain end to end.

\subsection{A Sign in Space}\label{sec:signin}

BL also participated in the 2023 \textit{A Sign in Space} project (de Paulis et al. 2023),
in which the ESA ExoMars Trace Gas Orbiter transmitted an encoded, simulated technosignature toward Earth.  The
transmission was received by multiple radio telescopes, including the BL-affiliated GBT, VLA, and ATA, providing a
coordinated validation of detection and follow-up workflows.

\subsection{FRBs}\label{sec:frbs}

The GBT backend's 4--8 GHz detection of FRB 121102 (Gajjar et al. 2018),
frequency-dependent-polarization capture of a bright FRB (Price et al. 2019), and multiband campaigns on repeating
sources (Sand et al. 2022) proved the raw-voltage chain end to end.

\subsection{Pulsars and the Galactic Center}\label{sec:pulsars}

Pulsar and magnetar programs doubled as commissioning science for new
modes, from 4--8 GHz GC magnetar spectroscopy (Suresh et al. 2021; Suresh et al. 2022) to a GBT redback-pulsar
discovery (Perez et al. 2023) and MeerKAT's commensal TRON detections (Section~\ref{sec:bluse}).

\subsection{Time-domain astrophysics on demand}\label{sec:timedomain}

BL-capable systems delivered radio follow-up of GRB 221009A (Bright
et al. 2023), the black-hole X-ray binary Swift J1727.8$-$1613 (Hughes et al. 2025), the fast blue optical transient
AT 2024wpp (Nayana et al. 2025), and tidal-disruption events (Sfaradi et al. 2025), evidence that SETI-grade wideband
backends are genuine observatory assets.

\subsection{Technosignature-adjacent commensal searches}\label{sec:adjacent}

The same Galactic Center data constrained axion-like particles (Foster et
al. 2022), underlining the value of open, reprocessable data.

\section{Future Facilities and the Road Ahead}\label{sec:future}

Looking ahead, the path from MeerKAT to SKA1-MID was anticipated in BL's early planning (Gajjar 2016;
Worden et al. 2017): phased arrays will make all-sky, always-on technosignature searching a byproduct of normal
observatory operations. BL has also participated in the ngVLA science case for SETI (Croft et al. 2018), and global
FRB-technosignature hybrids such as the CHIME/FRB Outriggers (CHIME/FRB Collaboration 2025) and GReX (Shila et al.
2025) embody the always-on transient infrastructure that a next-generation technosignature program can ride.  New projects in development expand the always-on operational model with a wide field of view.  The ALPACA receiver initially intended for the now-defunct Arecibo radio telescope is being deployed alongside the ATA without a dish for an all-sky field of view.  At Westerbork, the APERTIF feeds are being given new life by being pointed upward sans reflector to achieve a similarly wide instantaneous field of view.  Additionally, space- and lunar-based radio telescopes are being proposed and designed with technosignature searches as an explicit science goal.  The road ahead may indeed not be a road at all!

Across all of these future platforms, the design constants remain those of the commensal era: commensality, open
data, real-time classification, and multi-site coincidence.

\section{Lessons Learned and Conclusions}\label{sec:lessons}

\noindent
\textbf{Engineering:} Commodity hardware (Ethernet packet capture, COTS GPUs, software channelization) beat bespoke
hardware for SETI-scale raw capture, and every BL backend generation (GBT/ROACH2 $\rightarrow$ Parkes/ROACH-SNAP $\rightarrow$
COSMIC $\rightarrow$ ATA/Holoscan) pushed further in that direction. Raw-voltage archiving preserves optionality;
the archive repaid its storage cost with post-hoc searches (Painter et al. 2025; Pardo et al. 2025). Open data
multiplied scientific return far beyond the team's size.

\noindent
\textbf{Programmatic:} Observatory partnerships, built on commensal capability and data sharing, sustained access
that a private program could never have bought outright. Verification frameworks must exist \textit{before} the
signal of interest arrives; BLC1 set the standard. Target lists drive instruments: selection science (Isaacson et
al. 2017; Lacki et al. 2021) was the requirements document.

\noindent
\textbf{A decade in numbers:} Ten major facilities operationally engaged spanning $\sim$0.1--24+ GHz (plus the
SETI@home distributed-compute legacy); $\sim$42 PB of storage capacity deployed with $\sim$31 PB in use as of March
2026 (Table~\ref{tab:storage}); $>$1.2 million autonomously processed commensal pointings at MeerKAT since mid-2022
against a 32-million-star target list (Czech et al. 2026); and a program bibliography that already lists more than 200
papers.
Ten years in, the commensal era (streaming, GPU-native, ML-first, and open) is the default architecture for
technosignature science, and the SKA generation will inherit it.

\iacunnumbered{Acknowledgements}

Breakthrough Listen is managed by the Breakthrough Initiatives, sponsored by the Breakthrough Prize Foundation, and is
headquartered at the University of Oxford.  We are grateful to the staff and managements of our partner observatories:
the Green Bank Observatory is a facility of the National Science Foundation, operated under cooperative agreement by
Associated Universities, Inc.; the National Radio Astronomy Observatory is a facility of the National Science Foundation
operated under cooperative agreement by Associated Universities, Inc.; Murriyang and the Parkes Observatory are part of
the Australia Telescope National Facility, funded by the Australian Government for operation as a National Facility
managed by CSIRO; the MeerKAT telescope is operated by the South African Radio Astronomy Observatory, which is a
facility of the National Research Foundation, an agency of the Department of Science and Innovation; the Sardinia Radio
Telescope is operated by INAF and was funded by the Italian Department of Universities and Research (MIUR), the Italian
Space Agency (ASI), and the Autonomous Region of Sardinia; and the Allen Telescope Array refurbishment program and its
ongoing operations are substantially funded through the Franklin Antonio Bequest.  We acknowledge the Gamilaraay people
as the traditional owners of the land on which the Parkes Observatory stands.  Research at Lick Observatory is partially
supported by the Regents of the University of California.  We thank our Breakthrough Listen colleagues and the
engineering and operations teams whose dedication made these instruments and surveys possible.  Finally, we thank
everyone who has ever looked up at the night sky and wondered whether we are alone.

\begin{strip}
\iacunnumbered{Appendix A: Facility Inventory}
\centering
{\captionsetup{hypcap=false}%
\captionof{table}{Facility inventory}\label{tab:facilities}}
\fontsize{8}{9}\selectfont
\setlength{\tabcolsep}{3.5pt}
\begin{tabular}{@{}p{2.3cm}p{3.3cm}p{3.5cm}p{2.2cm}p{4.2cm}@{}}
\toprule
Facility & BL involvement & Backend / instrument & Freq. coverage & Key references \\
\midrule
GBT (100-m) & 2016--present & Wideband data recorder (CASPER ROACH2, packetized raw) & 0.7--8+ GHz & MacMahon+ 2018; Enriquez+ 2017; Price+ 2020; Gajjar+ 2018 \\
Parkes (64-m) & 2016--present & CASPER ROACH2/SNAP wideband recorders, 21-cm multibeam & 0.7--4 GHz & Price+ 2018; Choza+ 2024; Barrett+ 2025 \\
Automated Planet Finder & 2017--present (Gen 1) & Automated laser-search pipeline (optical spectrograph) & optical & Zuckerman+ 2023; Lipman+ 2019 \\
Keck/HIRES & 2016--2019 & High-res spectroscopy laser search & optical & Tellis \& Marcy 2017; Isaacson+ 2019 \\
MeerKAT & 2018--present (announced IAC Bremen 2018); BLUSE autonomous commensal since mid-2022 & Multicast USE ingest (full F-engine BW); BLUSE 64+1-beam survey; TRON interferometric mining & 544 MHz--3.5 GHz (UHF/L/S) & Czech+ 2021, 2026; Smirnov+ 2025a,b; Heywood+ 2024 \\
Sardinia Radio Telescope & 2019--present & High-frequency survey backend & $\sim$0.3--24 GHz & Pilia+ 2020; Manunza+ 2025 \\
LOFAR (intl. stations) & 2019--present & Dual-site coincidence search; post-processing barycentric correction & 110--190 MHz & Johnson+ 2023 \\
VERITAS & $\sim$2021--2023 & Imaging atmospheric Cherenkov (optical pulsed) & optical & Acharyya+ 2023 \\
VLA (COSMIC) & 2023--present & Ethernet commensal multimode backend, GPU cluster & 1--12+ GHz (extending) & Tremblay+ 2024, 2025; Stiegler+ 2025 \\
ATA (42-dish) & 2023--present (revival) & Wideband CASPER-based feeds/backends; Holoscan ML & 0.9--14+ GHz & Sheikh+ 2024; Tusay+ 2024; Joshi+ 2025; Ma+ 2025 \\
(Legacy) SETI@home & 1999--2020 & Volunteer distributed compute & n/a & Korpela+ 2025; Anderson+ 2025 \\
\bottomrule
\end{tabular}
\end{strip}

\iacunnumbered{Appendix B: Program Timeline}

\begin{itemize}
\item \textbf{2015}: Program launch (July 20, Royal Society); APF optical SETI begins (Dec 2015, 200 stars); GBT data recorder first light (Dec 2015)
\item \textbf{2016}: Parkes (Oct) Gen-1 recorders deployed
\item \textbf{2017}: First GBT survey papers; Keck and APF optical
\item \textbf{2018--2019}: FRB 121102 backend validation; archive \& formats paper (Lebofsky+ 2019); MeerKAT engagement announced (IAC Bremen 2018)
\item \textbf{2020--2021}: MeerKAT target selection; BLC1 episode; Exotica Catalog
\item \textbf{2022--2023}: Network years: LOFAR dual-site, SRT; solar gravitational lens attempt (Tusay+ 2022); BLUSE autonomous commensal survey begins at MeerKAT (mid-2022)
\item \textbf{2023--2024}: COSMIC commissioning on VLASS; ATA revival; deployed real-time ML
\item \textbf{2025--2026}: MeerKAT TRON commensal time-axis mining; BLUSE survey system paper (Czech+ 2026); archive-scale reprocessing; interstellar-object rapid response (3I/ATLAS); SKA-adjacent planning
\end{itemize}

\iacunnumbered{References}


\begin{enumerate}[label={[\arabic*]},leftmargin=1.8em]

\item Acharyya, A., et al. (2023). ``A VERITAS/Breakthrough Listen Search for Optical Technosignatures.'' \textit{AJ} 166(3):84. arXiv:2306.17680.
\item Anderson, D. P., et al. (2025). ``SETI@home: Data Analysis and Findings.'' \textit{AJ} 170(2):111. arXiv:2506.14737.
\item Barrett, R. A. W., et al. (2025). ``Breakthrough Listen: A technosignature search around 27 eclipsing exoplanets selected from TESS.'' \textit{PASA} 42:e089. arXiv:2506.13459.
\item Bright, J. S., et al. (2023). ``Precise measurements of self-absorbed rising reverse shock emission from gamma-ray burst 221009A.'' \textit{Nature Astronomy} 7:986--995. arXiv:2303.13583.
\item Brzycki, B., et al. (2020). ``Narrow-band Signal Localization for SETI on Noisy Synthetic Spectrogram Data.'' \textit{PASP} 132(1017):114501. arXiv:2006.04362.
\item Brzycki, B., et al. (2022). ``Setigen: Simulating Radio Technosignatures for the Search for Extraterrestrial Intelligence.'' \textit{AJ} 163(5):222. arXiv:2203.09668.
\item CHIME/FRB Collaboration (2025). ``CHIME/FRB Outriggers: Design Overview.'' \textit{ApJ} 993(1):55. arXiv:2504.05192.
\item Choza, C., et al. (2024). ``The Breakthrough Listen Search for Intelligent Life: Technosignature Search of 97 Nearby Galaxies.'' \textit{AJ} 167(1):10. arXiv:2312.03943.
\item Croft, S., et al. (2018). ``Science with an ngVLA: SETI Searches for Evidence of Intelligent Life in the Galaxy.'' \textit{ngVLA Science Book}. arXiv:1810.06568.
\item Croft, S., Siemion, A., and Hilburn, B. (2019). ``Open Sourcing the Search for Extraterrestrial Intelligence.'' \textit{GRCon2019}.
\item Croft, S. and Pollak, A. (2020). ``Are We Alone? How GNU Radio Can Help Us Find ET.'' \textit{GR-Con2020}.
\item Czech, D., et al. (2021). ``The Breakthrough Listen Search for Intelligent Life: MeerKAT Target Selection.'' \textit{PASP} 133(1024):064502. arXiv:2103.16250.
\item Czech, D. J., et al. (2026). ``Breakthrough Listen's Automated Commensal Technosignature Survey with MeerKAT.'' \textit{MNRAS} (in preparation).
\item Davenport, J. R. A., et al. (2022). ``Searching the SETI Ellipsoid with Gaia.'' \textit{AJ} 164(3):117. arXiv:2206.04092.
\item Davenport, J. R. A., et al. (2023). ``Real-Time Technosignature Strategies with SN 2023ixf.'' arXiv:2306.03118.
\item de Paulis, D., et al. (2023). ``A sign in space: An interdisciplinary exploration of the potential reception of an extraterrestrial signal.'' \textit{Acta Astronautica} 212:436--440. doi:10.1016/j.actaastro.2023.08.026.
\item Enriquez, J. E., et al. (2017). ``The Breakthrough Listen Search for Intelligent Life: 1.1--1.9 GHz Observations of 692 Nearby Stars.'' \textit{ApJ} 849:104. arXiv:1709.03491.
\item Farah, W., et al. (2024). ``Very Bright Radio Flare Observed from Cyg X-3 with the ATA.'' \textit{The Astronomer's Telegram} 16466.
\item Foster, J. W., et al. (2022). ``Extraterrestrial Axion Search with the Breakthrough Listen Galactic Center Survey.'' \textit{Phys. Rev. Lett.} 129(25):251102. arXiv:2202.08274.
\item Franz, N., et al. (2022). ``The Breakthrough Listen Search for Intelligent Life: Technosignature Search of Transiting TESS Targets of Interest.'' \textit{AJ} 163(3):104. arXiv:2201.00918.
\item Gallay, E. M., et al. (2025). ``Technosignature Searches with Real-time Alert Brokers.'' \textit{AJ} 170(2):95. arXiv:2506.14744.
\item Gajjar, V. (2016). ``Commensal SETI in the SKA era.'' \textit{SKA2016: Science for the SKA Generation}.
\item Gajjar, V., et al. (2018). ``Highest Frequency Detection of FRB 121102 at 4--8 GHz Using the Breakthrough Listen Digital Backend at the Green Bank Telescope.'' \textit{ApJ} 863:2. arXiv:1804.04101.
\item Gajjar, V., et al. (2021). ``The Breakthrough Listen Search For Intelligent Life Near the Galactic Center. I.'' \textit{AJ} 162(1):33. arXiv:2104.14148.
\item Heywood, I., et al. (2024). ``MeerKAT Observations of Procyon at 815.5 MHz.'' \textit{RNAAS} 8(10):255. arXiv:2410.15069.
\item Houston, K., Siemion, A., and Croft, S. (2021). ``Strategies for Maximizing Detection Rate in Radio SETI.'' \textit{AJ} 162(4):151. arXiv:2106.06594.
\item Hughes, A. K., et al. (2025). ``Comprehensive Radio Monitoring of the Black Hole X-Ray Binary Swift J1727.8$-$1613 during Its 2023--2024 Outburst.'' \textit{ApJ} 988(1):109. arXiv:2506.07798.
\item Isaacson, H., et al. (2017). ``The Breakthrough Listen Search for Intelligent Life: Target Selection of Nearby Stars and Galaxies.'' \textit{PASP} 129(5):054501. arXiv:1701.06227.
\item Isaacson, H., et al. (2019). ``The Breakthrough Listen Search for Intelligent Life: No Evidence of Claimed Periodic Spectral Modulations in High-Resolution Optical Spectra of Nearby Stars.'' \textit{PASP} 131(995):014201. arXiv:1812.02258.
\item Jacobson-Bell, B., et al. (2025a). ``Anomaly Detection and Radio-frequency Interference Classification with Unsupervised Learning in Narrowband Radio Technosignature Searches.'' \textit{AJ} 169(4):206. arXiv:2411.16556.
\item Jacobson-Bell, B., et al. (2025b). ``Breakthrough Listen Observations of 3I/ATLAS with the Green Bank Telescope at 1--12 GHz.'' \textit{RNAAS} 9(12):351.
\item Johnson, O. A., et al. (2023). ``A Simultaneous Dual-site Technosignature Search Using International LOFAR Stations.'' \textit{AJ} 166(5):193. arXiv:2310.15704.
\item Joshi, P., et al. (2025). ``High Frequency Wideband Study of FRB20240114A with the Allen Telescope Array.'' \textit{43rd ASI Meeting} P60.
\item Korpela, E. J., et al. (2025). ``SETI@home: Data Acquisition and Front-end Processing.'' \textit{AJ} 170(2):112. arXiv:2506.14718.
\item Lacki, B. C., et al. (2021). ``One of Everything: The Breakthrough Listen Exotica Catalog.'' \textit{ApJS} 257(2):42. arXiv:2006.11304.
\item Lebofsky, M., et al. (2019). ``The Breakthrough Listen Search for Intelligent Life: Public Data, Formats, Reduction, and Archiving.'' \textit{PASP} 131(1006):124505. arXiv:1906.07391.
\item Lipman, D., et al. (2019). ``The Breakthrough Listen Search for Intelligent Life: Searching Boyajian's Star for Laser Line Emission.'' \textit{PASP} 131:034202. arXiv:1812.10161.
\item Ma, P. X., et al. (2023). ``A deep-learning search for technosignatures from 820 nearby stars.'' \textit{Nature Astronomy} 7:492--502.
\item Ma, P. X., et al. (2024). ``A deep neural network based reverse radio spectrogram search algorithm.'' \textit{RAS Techniques and Instruments} 3(1):33--43. arXiv:2302.13854.
\item Ma, P. X., et al. (2025). ``A deployed real-time end-to-end deep learning algorithm for fast radio burst detection.'' \textit{A\&A} 702:A181. doi:10.1051/0004-6361/202555217.
\item MacMahon, D. H. E., et al. (2018). ``The Breakthrough Listen Search for Intelligent Life: A Wideband Data Recorder System for the Robert C. Byrd Green Bank Telescope.'' \textit{PASP} 130(4):044502. arXiv:1707.06024.
\item Manunza, L., et al. (2025). ``The first high frequency technosignature search survey with the Sardinia Radio Telescope.'' \textit{Acta Astronautica} 233:155--167. arXiv:2410.09288.
\item Murphy, C., Croft, S., Dana, R., and DeBoer, D. (2019). ``The Breakthrough Listen Search for Intelligent Life: Characterization and Mitigation of Satellite RFI.'' \textit{AAS Meeting Abstracts} 233, \#363.12.
\item Nayana, A. J., et al. (2025). ``The Most Luminous Known Fast Blue Optical Transient AT 2024wpp: Unprecedented Evolution and Properties in the X-Rays and Radio.'' \textit{ApJ} 993(1):L6. arXiv:2509.00952.
\item Nilipour, A., et al. (2023). ``Signal Synchronization Strategies and Time Domain SETI with Gaia DR3.'' \textit{AJ} 166(2):79. arXiv:2308.00066.
\item Pardo, S., et al. (2025). ``Using Anomaly Detection to Search for Technosignatures in Breakthrough Listen Observations.'' \textit{AJ} 170(1):12. arXiv:2505.03927.
\item Perez, K. I., et al. (2022). ``Breakthrough Listen Search for the WOW! Signal.'' \textit{RNAAS} 6(9):197.
\item Perez, K. I., et al. (2023). ``Green Bank Telescope Discovery of the Redback Binary Millisecond Pulsar PSR J0212+5321.'' \textit{ApJ} 952(2):150. arXiv:2306.04951.
\item Pilia, M., et al. (2020). ``The Lowest-frequency Fast Radio Bursts: Sardinia Radio Telescope Detection of the Periodic FRB 180916 at 328 MHz.'' \textit{ApJ} 896(2):L40. arXiv:2003.12748.
\item Price, D. C. (2016). ``Breakthrough Listen data formats and compute framework.'' \textit{GBT SETI Workshop}, Green Bank WV.
\item Price, D. C., et al. (2018). ``The Breakthrough Listen Search for Intelligent Life: Wide-bandwidth digital instrumentation for the CSIRO Parkes 64-m telescope.'' \textit{PASA} 35:e041. arXiv:1804.04571.
\item Price, D. C., et al. (2019). ``A fast radio burst with frequency-dependent polarization detected during Breakthrough Listen observations.'' \textit{MNRAS} 486(3):3636--3646. arXiv:1901.07412.
\item Price, D. C., et al. (2020). ``The Breakthrough Listen Search for Intelligent Life: Observations of 1327 Nearby Stars Over 1.10--3.45 GHz.'' \textit{AJ} 159(3):86. arXiv:1906.07750.
\item Price, D. C. (2024). ``Reduced-resolution beamforming: Lowering the computational cost for pulsar and technosignature surveys.'' \textit{PASA} 41:e037. arXiv:2402.12723.
\item Sand, K. R., et al. (2022). ``Multiband Detection of Repeating FRB 20180916B.'' \textit{ApJ} 932(2):98. arXiv:2111.02382.
\item Sagan, C. (1979). \textit{Broca's Brain: The Romance of Science}. Random House.
\item Sheikh, S. Z., et al. (2019). ``Choosing a Maximum Drift Rate in a SETI Search: Astrophysical Considerations.'' \textit{ApJ} 884(1):14. arXiv:1910.01148.
\item Sheikh, S. Z., et al. (2020). ``The Breakthrough Listen Search for Intelligent Life: A 3.95--8.00 GHz Search for Radio Technosignatures in the Restricted Earth Transit Zone.'' \textit{AJ} 160(1):29. arXiv:2002.06162.
\item Sheikh, S. Z., et al. (2021a). ``Analysis of the Breakthrough Listen signal of interest blc1 with a technosignature verification framework.'' \textit{Nature Astronomy} 5:1153--1162. arXiv:2111.06350.
\item Sheikh, S. Z., et al. (2021b). ``No Redetections of blc1 in 39 hr of Reobservation Campaigns of Proxima Centauri.'' \textit{RNAAS} 5(10):248.
\item Sheikh, S. Z., et al. (2023). ``A Green Bank Telescope Search for Narrowband Technosignatures between 1.1 and 1.9 GHz During 12 Kepler Planetary Transits.'' \textit{AJ} 165(2):61. arXiv:2212.05137.
\item Sheikh, S. Z., et al. (2024). ``Characterization of the repeating FRB 20220912A with the Allen Telescope Array.'' \textit{MNRAS} 527(4):10425--10439. arXiv:2312.07756.
\item Sheikh, S. Z., et al. (2025). ``A Search for Radio Technosignatures from Interstellar Object 3I/ATLAS with the Allen Telescope Array.'' arXiv:2512.18142.
\item Shila, K. A., et al. (2025). ``GReX: An Instrument Overview and New Upper Limits on the Galactic FRB Population.'' \textit{PASP} 137(7):075001. arXiv:2504.18680.
\item Sfaradi, I., et al. (2025). ``Radio upper limits from the TDE 2025aarm using the Allen Telescope Array.'' \textit{Transient Name Server AstroNote} 313.
\item Smirnov, O. M., et al. (2025a). ``Mining the time axis with TRON -- I. Millisecond pulsars in Omega Centauri, Terzan 5, and 47 Tucanae detected through MeerKAT interferometric imaging.'' \textit{MNRAS} 538(1):L62--L68. arXiv:2501.09488.
\item Smirnov, O. M., et al. (2025b). ``Mining the time axis with TRON -- II. MeerKAT detects a stellar radio flare from a distant RS CVn candidate.'' \textit{MNRAS} 538(1):L89--L93. arXiv:2501.09489.
\item Smith, S., et al. (2021). ``A radio technosignature search towards Proxima Centauri resulting in a signal of interest.'' \textit{Nature Astronomy} 5:1148--1152. arXiv:2111.08007.
\item Sofair, J., et al. (2024). ``RFI Flagging with Excess Kurtosis in the COSMIC SETI Technosignature Post-Processing Pipeline.'' \textit{AAS Meeting Abstracts} 243, \#159.03.
\item Steffes, L., et al. (2024). ``ARTISTIC: A New Pipeline for Radio Extraterrestrial Technosignature Searches using COSMIC.'' \textit{AAS Meeting Abstracts} 243, \#109.01.
\item Stiegler, N., et al. (2025). ``A High Frequency Technosignature Search with COSMIC and the VLA.'' \textit{AAS Meeting Abstracts} 245, \#363.01.
\item Suresh, A., et al. (2021). ``4--8 GHz Spectrotemporal Emission from the Galactic Center Magnetar PSR J1745$-$2900.'' \textit{ApJ} 921(2):101. arXiv:2108.05404.
\item Suresh, A., et al. (2022). ``4--8 GHz Fourier-domain Searches for Galactic Center Pulsars.'' \textit{ApJ} 933(2):121. arXiv:2203.00036.
\item Tellis, N. K. and Marcy, G. W. (2017). ``A Search for Laser Emission with Megawatt Thresholds from 5600 FGKM Stars.'' \textit{AJ} 153:251. arXiv:1704.02535.
\item Traas, R., et al. (2021). ``The Breakthrough Listen Search for Intelligent Life: Searching for Technosignatures in Observations of TESS Targets of Interest.'' \textit{AJ} 161(6):286. arXiv:2101.11137.
\item Tremblay, C. D., et al. (2024). ``COSMIC: An Ethernet-based Commensal, Multimode Digital Backend on the Karl G. Jansky Very Large Array for the Search for Extraterrestrial Intelligence.'' \textit{AJ} 167(1):35. arXiv:2310.09414.
\item Tremblay, C. D., et al. (2025). ``COSMIC's Large-scale Search for Technosignatures during the VLA Sky Survey: Survey Description and First Results.'' \textit{AJ} 169(3):122. arXiv:2501.17997.
\item Truzzi, M. (1978). ``On the Extraordinary: An Attempt at Clarification.'' \textit{Zetetic Scholar} 1(1):11.
\item Tusay, N., et al. (2022). ``A Search for Radio Technosignatures at the Solar Gravitational Lens Targeting Alpha Centauri.'' \textit{AJ} 164(3):116. arXiv:2206.14807.
\item Tusay, N., et al. (2024). ``A Radio Technosignature Search of TRAPPIST-1 with the Allen Telescope Array.'' \textit{AJ} 168(6):283. arXiv:2409.08313.
\item Wandia, K., et al. (2023). ``An interferometric SETI observation of Kepler-111 b.'' \textit{MNRAS} 522(3):3784--3794. arXiv:2305.02262.
\item Worden, S. P., et al. (2017). ``Breakthrough Listen -- a new search for life in the universe.'' \textit{Acta Astronautica} 139:98--101. doi:10.1016/j.actaastro.2017.06.008.
\item Wright, J. T., et al. (2017). ``Green Bank 100 m telescope observations of Boyajian's Star from 1--27.5 GHz with the Breakthrough Listen backend.'' \textit{NASA Astrobiology Science Conference}, poster \#3449.
\item Zhang, Y. G., et al. (2018a). ``Fast radio burst 121102 pulse detection and periodicity: A machine learning approach.'' \textit{ApJ} 866(2):149.
\item Zhang, Y. G., et al. (2018b). ``Self-supervised anomaly detection for narrowband seti.'' \textit{IEEE GlobalSIP}:1114--1118. arXiv:1901.04636.
\item Zuckerman, A., et al. (2023). ``The Breakthrough Listen Search for Intelligent Life: A Laser Search Pipeline for the Automated Planet Finder.'' \textit{AJ} 165(3):114. arXiv:2301.06971.

\end{enumerate}


\end{document}